\documentclass{iopart}
\usepackage{iopams}  
\usepackage{graphicx}
\def\be{\begin{equation}}
\def\ee{\end{equation}}

\begin{document}

\title{ 
Lattice calculations of meson correlators 
and spectral functions at finite temperature
}

\author{P\'eter Petreczky 
\footnote[1]{
Goldhaber Fellow\\
This work has been authored under contract number DE-AC02-98CH10886 with 
the U.S. Department of Energy
}
}
\address{ 
Nuclear Theory Group, Department of Physics, Brookhaven
National Laboratory, Upton, New York, 11973\\
email:	petreczk@bnl.gov
}

\begin{abstract}
I review recent progress in relating meson spectral
function to imaginary time correlation function at finite temperature
calculated on isotropic as well as on anisotropic
lattices. Special attention is payed for the lattice artifacts
present in calculation of meson spectral functions.
Results in the case of light quarks as well as heavy
quarks are reviewed which indicate in particular that
even in the chiral limit meson spectral functions have non-trivial
structure and the ground state quarkonia survive up to
temperature $1.5T_c$.
\end{abstract}


\submitto{\JPG}


\section{Introduction}
Spectral functions of mesonic operators play an important role
in finite temperature QCD. Many experimental results
in high energy heavy ion collisions ( e.g. low invariant mass
dilepton enhancement, anomalous $J/\psi$ suppression etc.)
can be understood in terms of medium modifications of meson
spectral functions \cite{rapp}.

It is commonly believed that lattice QCD is capable only for calculation
of static quantities at finite temperatures, such as the transition temperature,
equation of state, screening lengths etc. However, it was shown by
Asakawa, Hatsuda and Nakahara that using the {\em Maximum Entropy Method}
one can in principle reconstruct also meson spectral functions.
The method was successfully applied at zero temperature \cite{asakawa01,cppacs}
and later also at finite temperature \cite{karsch02,wetzorke02,karsch03,
datta03,asakawa02,umeda2,wetzorke03}. Though systematic uncertainties in 
the spectral function calculated on lattice are not yet completely 
understood, it was shown in Ref. \cite{karsch02} that precise determination 
of the imaginary time correlator can alone provide stringent constraints 
on the spectral function at finite temperature. 

In this contribution I am going to review recent results on
meson correlators and spectral functions in finite temperature QCD.
The rest of the paper is organized as follows. In section 2 I will
discuss the relation between the imaginary time correlators and
meson spectral functions. In section 3 numerical results for the light
quarks will be discussed. Section 4 contains results on charmonia
at finite temperature. Finally conclusions are given in section 5.

\section{Meson correlators and spectral function in continuum
and lattice QCD}

In finite temperature field theory one usually considers two
types of real time correlation function 
of some operator $\hat O$ 
\be
D^{>}(t,t')=\langle \hat O(t) \hat O(t') \rangle_T,~~~
D^{<}(t,t')=\langle \hat O(t') \hat O(t) \rangle_T,
\ee
where $\langle ... \rangle_T = \langle ... e^{-\hat{H}/T} 
\rangle_T$ denotes the thermal average \cite{lebellac}.
The imaginary time correlation function which can be calculated also
by using lattice simulations is defined as simple analytic 
continuation 
\be
G(\tau)=\langle {\cal T} \hat O(-i \tau) \hat O(0) \rangle.
\label{rel1}
\ee
( ${\cal T}$ stands for time ordered product).
The spectral function can be defined through the Fourier transform of
$D^{>(<)}(t)$ as
\be
\displaystyle
\sigma(\omega)=\frac{D^{>}(\omega)-D^{<}(\omega)}{2 \pi}=\frac{1}{\pi} Im D_R(\omega),
\ee
where $D_R(\omega)$ is the retarded correlator.
Using Eq. (\ref{rel1}) and KMS condition on $D^{>(<)}$ \cite{lebellac}
one can easily derive the following integral relation between the 
imaginary time correlator and the spectral function
\be
\displaystyle
G(\tau)=\int_0^{\infty} d \omega \sigma(\omega) \frac{\cosh(\omega (\tau-1/(2T))}
{\sinh\frac{\omega}{2T}} \equiv \int_0^{\infty} d \omega \sigma(\omega) K(\omega,\tau)
\label{rel2}
\ee
Using this relation one can in principle reconstruct the spectral function
by calculating $G(\tau)$ on lattice.
At zero temperature the kernel, $K(\omega,\tau)$ reduces to simple
exponential and at large Euclidean times the correlation function 
picks up the contribution from the lowest lying meson state in $\sigma(\omega)$,
i.e $G(\tau)=\exp(-m \tau)$. At finite temperature the analysis of the 
asymptotic behavior of the correlators is no longer possible as $\tau$ is limited
to the interval $[0,~1/T]$  where excited states are equally important as the
ground state. Additional complications arise in lattice calculations where
correlators are calculated only at finite set of Euclidean times
$\tau T=k/N_{\tau}$, $k=0,...,N_{\tau}-1$  
with $N_{\tau}$   being the temporal extent of the lattice.
In order to reconstruct the spectral functions from this 
limited information it is necessary to include in the statistical
analysis of the numerical results also prior information on the 
structure of $\sigma(\omega)$ (e.g. such as $\sigma(\omega)>0$ for
$\omega>0$). This can be done through the application of the
{\em Maximum Entropy Method} (MEM) \footnote{Other methods of
introducing prior information into the statistical analysis have
been also discussed in Ref. \cite{lepage} for the zero temperature
case and in Ref. \cite{gupta-v} for finite temperature QCD.}.

In order to study meson properties at finite temperature appropriate choice
of the operator $\hat O$ should be made. One possible choice is local
meson operator bilinear in quark-antiquark fields (current)
\cite{asakawa01,cppacs,karsch02,wetzorke02}
\be
O_H(\tau,\vec{x})=Z_H \bar q(\tau,\vec{x}) \Gamma_H q(\tau,\vec{x}),
~~~\Gamma_H=1, \gamma_5, \gamma_{\mu},
\gamma_5 \gamma_{\mu}
\label{pointop}
\ee
for scalar, pseudoscalar, vector and axial vector channels correspondingly.
The normalization constant $Z_{H}$ relates the current calculated
in lattice regularization scheme to current in $\overline{MS}$ scheme with
$\mu_{\overline{MS}}=a^{-1}$ (with $a$ being the lattice spacing).
We can define then the temporal correlators at finite spatial momentum
$\vec{p}$ 
\be
G_{H}(\tau,\vec{p})=\langle O_H(\tau,\vec{p}) O_H^{\dagger}(\tau,-\vec{p}) \rangle,
O(\tau,\vec{p})=\sum_{\vec{x}} e^{i \vec{p} \vec{x}} O_H(\tau,\vec{x})
\ee
One can also consider correlators of extended operators defined as \cite{umeda2,umeda1}
\be
\tilde O_H(\tau,\vec{x})=\sum_{\vec{y}} \phi(\vec{y}) \bar q(\tau,\vec{x}) \Gamma
q(\tau,\vec{x}+\vec{y}).
\ee
Here 
\be
\phi(\vec{y})=\exp(-b |\vec{y}|^p)
\label{triwf}
\ee
is the trial wave function
which controls the size of the meson source and can be regarded as a 
physical input to the problem. The use of extended operators makes
the reconstruction of the spectral function easier as the correlator is
dominated by a single peak in the spectral function \cite{umeda2}.
However, the corresponding spectral function is not related to a physically
observable quantity, it can provide information only about the mass
and the width of the resonance but not about the corresponding decay constant.
This also only works for sharp resonances.
At high temperature "mesons" will appear as broad structures in the
spectral function and the above method is no longer applicable.

Though the in-medium properties of mesons are encoded primarily in the 
temporal correlator and the corresponding spectral function,
in finite temperature QCD it is customary to study spatial correlators
defined as
\be
G_{H}=\langle O(z) O^{\dagger}(0) \rangle,~~~
O(z)=\sum_{\tau,x,y} O(\tau,x,y,z).
\ee
Since the lattice extent is not limited in the spatial directions 
one can study their large distance behavior. At large distances
the spatial correlators decay exponentially and the exponential 
decay governed by the so-called screening mass. The spatial
correlators can be related to the spectral function via 
following relation
\be
\hspace*{-2.1cm}
G_{H}(z)=\int_{-\infty}^{\infty} d p_z e^{i p_z z}
\int_0^{1/T} G(\tau,0,0,p_z)=\int_{-\infty}^{\infty} d p_z e^{i p_z z}
\int_0^{\infty} d \omega \frac{\sigma(\omega,0,0,p_z)}{\omega}
\ee
In general the screening masses are different from the pole
masses, however, from the above equation one can easily see
that in the special case when the spectral function is dominated by a single 
$\delta$-function for small $\omega$ the screening and the pole
masses are equal.

\begin{figure}
\hspace*{-1cm}
\includegraphics[width=7cm]{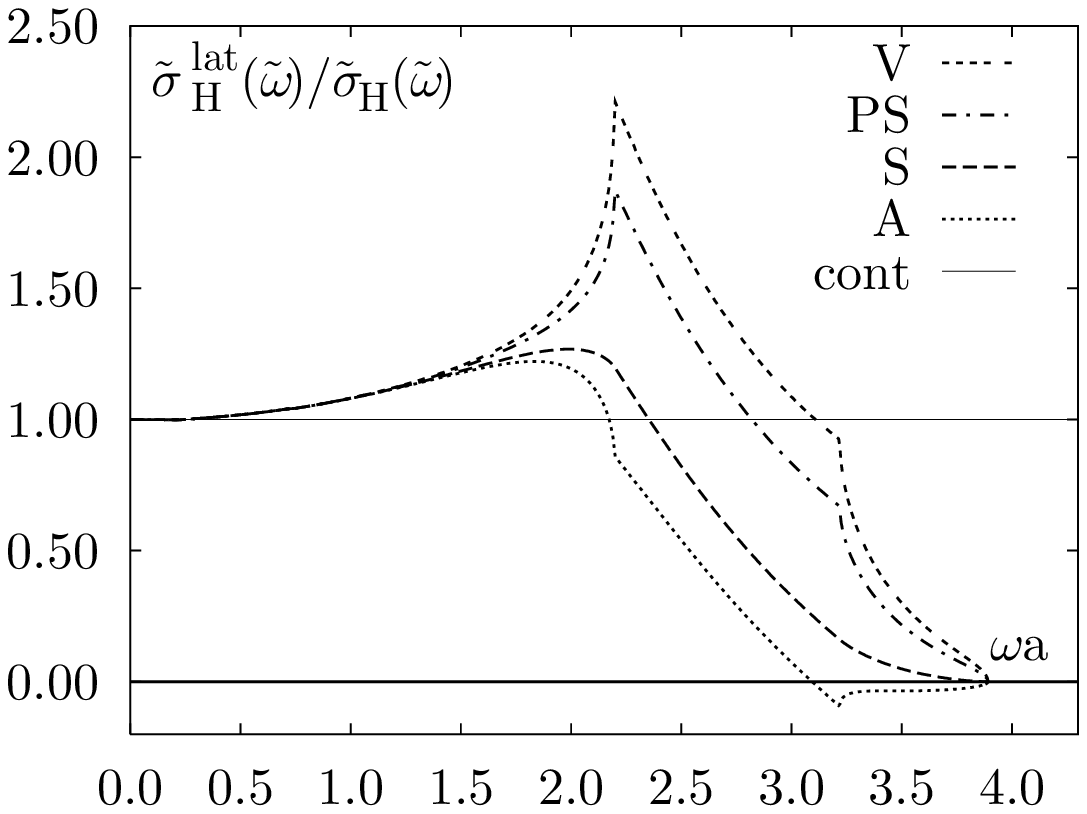}
\includegraphics[width=7cm]{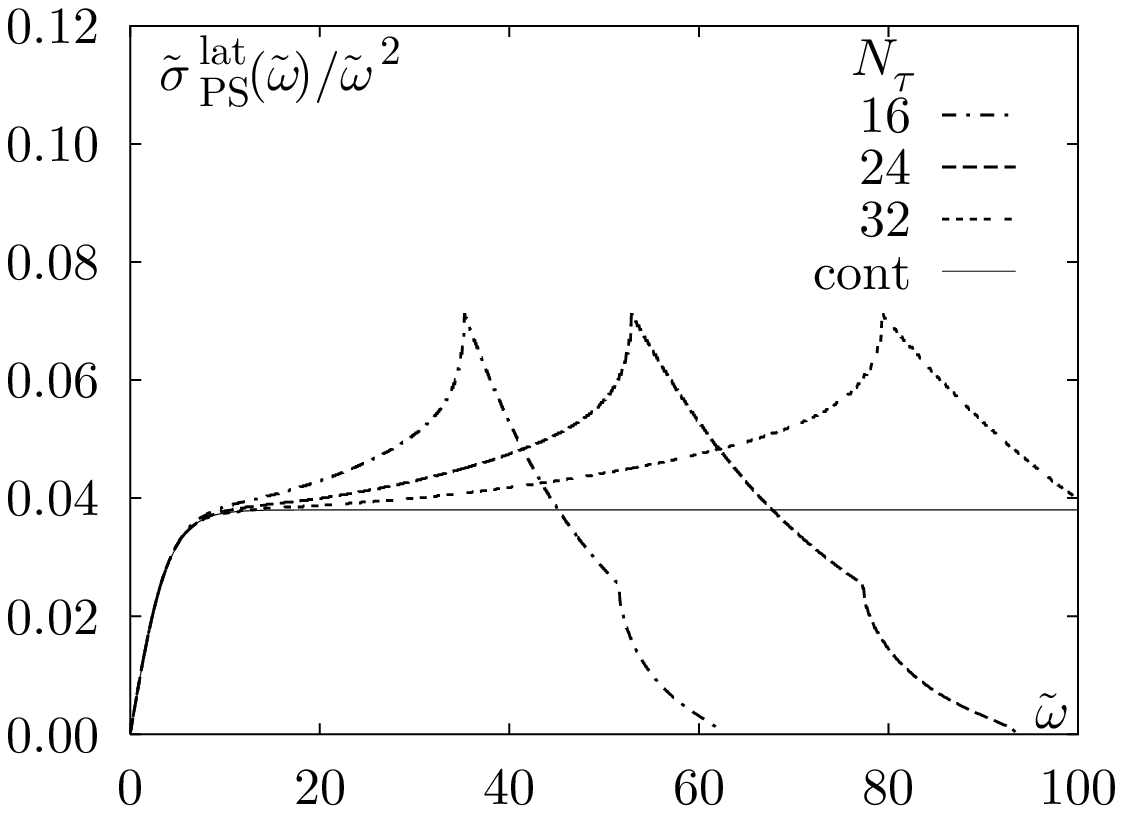}
\caption{The ratio of the lattice and continuum spectral functions
in different channels (left) versus $\omega a$ and the lattice pseudoscalar
spectral function versus $\tilde \omega=\omega/T$ for different
values of $N_{\tau}$ calculated using Wilson fermions.
See Ref. \cite{karsch03a} for further details.}
\label{latspf}
\end{figure}
The relation between the imaginary time correlator and 
the spectral function (\ref{rel2}) is valid only in the continuum theory. 
As the correlators are calculated on lattice the question arises 
whether the spectral representation of the form (\ref{rel2}) can
be derived also for the lattice correlator.
In ref. \cite{karsch03a} it was shown that this is indeed the case
in the free theory and that the cutoff effects present in the 
correlator at small separations are contained in the spectral
function. Because of the asymptotic freedom calculation in the free
theory are relevant at high temperature and/or large energies $\omega$.
In Fig. \ref{latspf} I show the ratio of of the lattice spectral function 
to the corresponding continuum spectral function versus $a \omega$
as well as the lattice pseudoscalar spectral function versus
$\omega/T$ in the free theory. As one can see from the figures the
 lattice spectral functions start to deviate from the continuum one for 
$\omega a>0.5$  and in the pseudoscalar and vector channels show
a peak like structure around $\omega a \simeq 2$. The lattice 
spectral functions vanishes for $\omega a > \omega_{max} a \sim 4 N_{\tau}$.
As the temporal extent of the lattice $N_{\tau}$ is increased for 
fixed temperature, i.e. the lattice spacing decreases ($a=1/(N_{\tau} T$))
the peak structure moves to larger value of $\omega/T$ (see Fig. \ref{latspf}).
Peak structure around $\omega a \simeq 2$ was also observed in the
interacting theory \cite{cppacs}. 
The main reason for the large cutoff effects in the lattice spectral function is
that on lattice the quark dispersion relation gets modified 
at large momenta, $pa >1$, relative to its continuum counterpart.
Cutoff effects in the spectral
function can be substantially reduced by using an improved fermion action,
the so-called truncated perfect action on hypercube \cite{biet96}
for which the lattice 
quark dispersion relation is much closer to the continuum even for 
$pa >1$ \cite{karsch03a}.

\section{Numerical results in the light quark sector}

Meson correlators and spectral functions for small 
quark masses were intensively studied during recent 
years \cite{karsch02,wetzorke02,karsch03,qcdtaro}.
In Refs. \cite{karsch02,wetzorke02,karsch03} 
spectral functions were studied using non-perturbatively
${\cal O}(a)$ improved Wilson action \cite{luescher} 
and isotropic $32^3 \times 16$,
$48^3 \times 12$, $64^3 \times 16$ and $64^3 \times 24$ lattices.
In these studies quark masses corresponding to the pion masses
in the range 400MeV to 1Gev were considered in the confined phase,
while in the deconfined phase calculation were performed in the
chiral limit and the temperature interval was $0.4T_c-3T_c$ with
$T_c$ being the deconfinement temperature. The renormalization
factors $Z_{H}$ appearing in Eq. \ref{pointop} were
determined non-perturbatively in Ref. \cite{luescher} for
the vector and axial vector channels, while for the scalar
and pseudoscalar channels they were calculated using tadpole
improved perturbation theory.
\begin{figure}
\includegraphics[width=8cm]{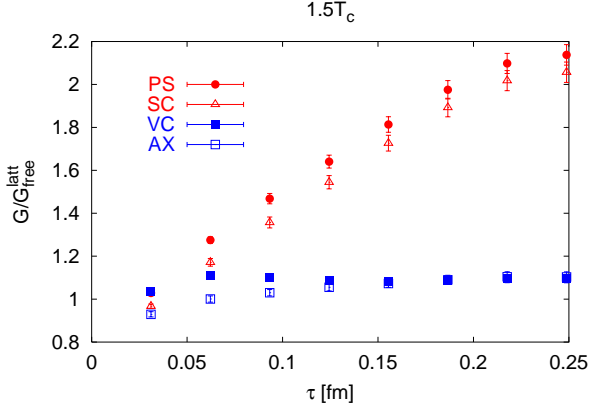}
\caption{The ratio of meson correlators at $1.5T_c$ in
different channels to the corresponding correlators in the
free theory calculated on $64^3 \times 16$ lattice.}
\label{restor}
\end{figure}
Asakawa, Hatsuda and Nakahara have studied meson spectral functions 
using $32^3 \times N_{\tau}$ anisotropic lattices with 
$a_{\tau}/a_{\sigma}=4$ and $N_{\tau}=96-32$ corresponding to temperatures
$0.8T_c-3T_c$ \cite{asakawa02}. The quark masses used in this 
study correspond to $m_{\pi}/m_{\rho}=0.7$. 
Below the deconfinement temperature neither the correlators nor
the spectral functions indicate in-medium change of meson properties
contrary to existing theoretical predictions \cite{rapp}. This
is likely due to the quenched approximation and large quark masses
used in these studies.
\begin{figure}
\hspace*{-1cm}
\includegraphics[width=7cm]{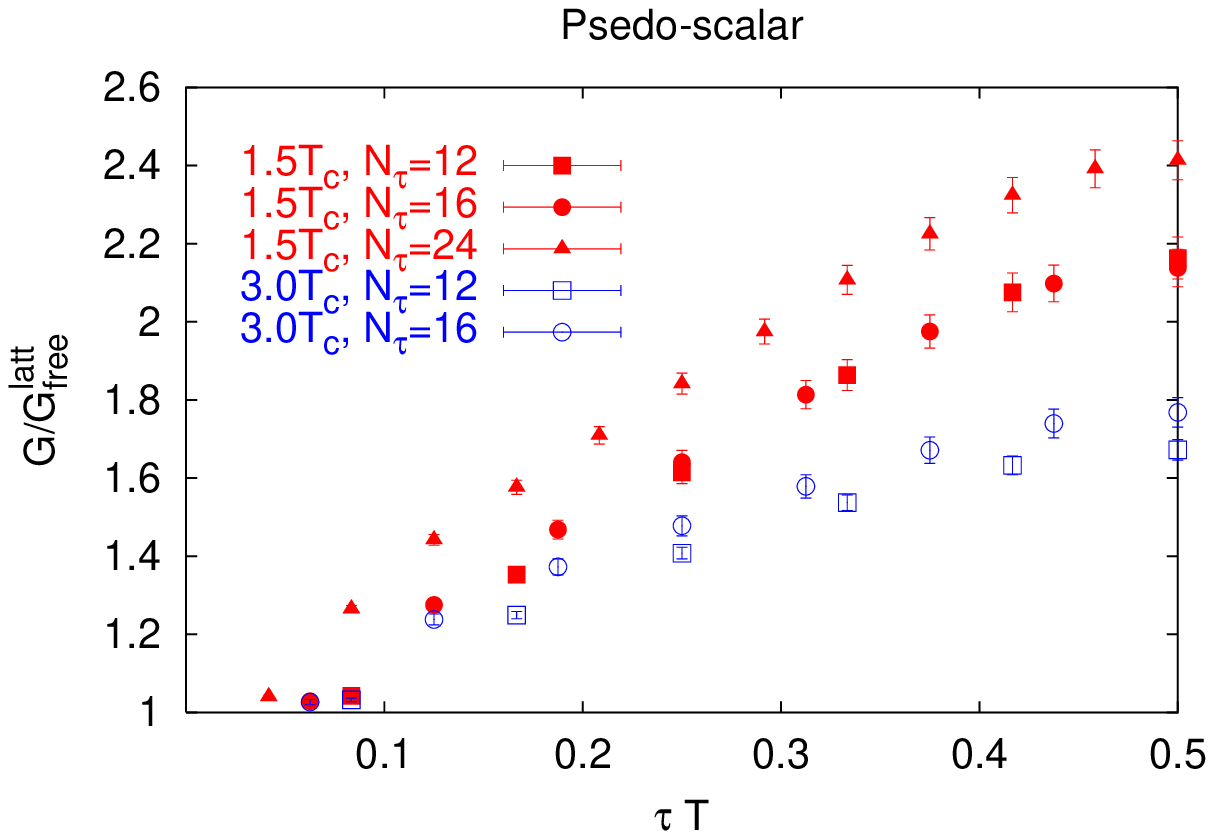}
\includegraphics[width=7cm]{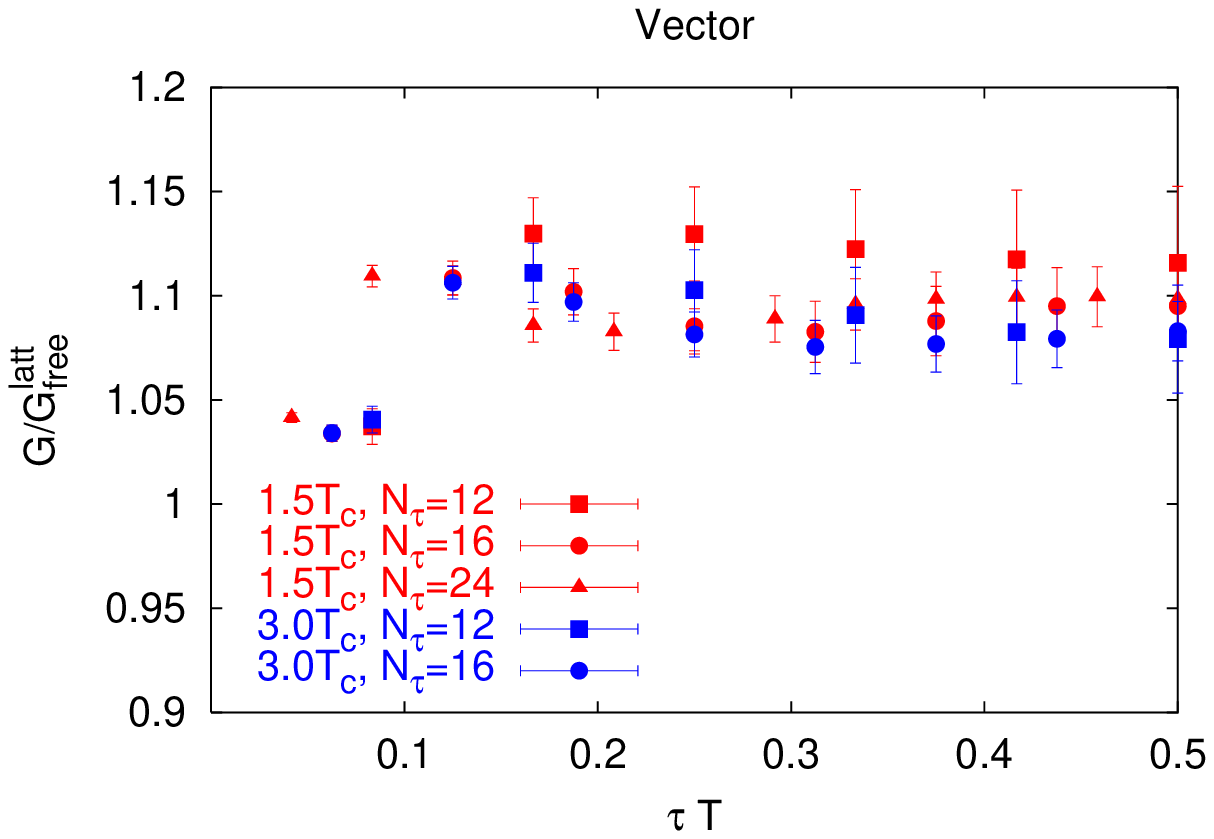}
\caption{The ratio of meson correlators at $1.5T_c$ and $3T_c$ in
the pseudoscalar (left) and vector (right) channels 
to the corresponding correlators in the
free theory calculated on lattice.}
\label{corpsvc}
\end{figure}

\begin{figure}
\hspace*{-1cm}
\includegraphics[width=7cm]{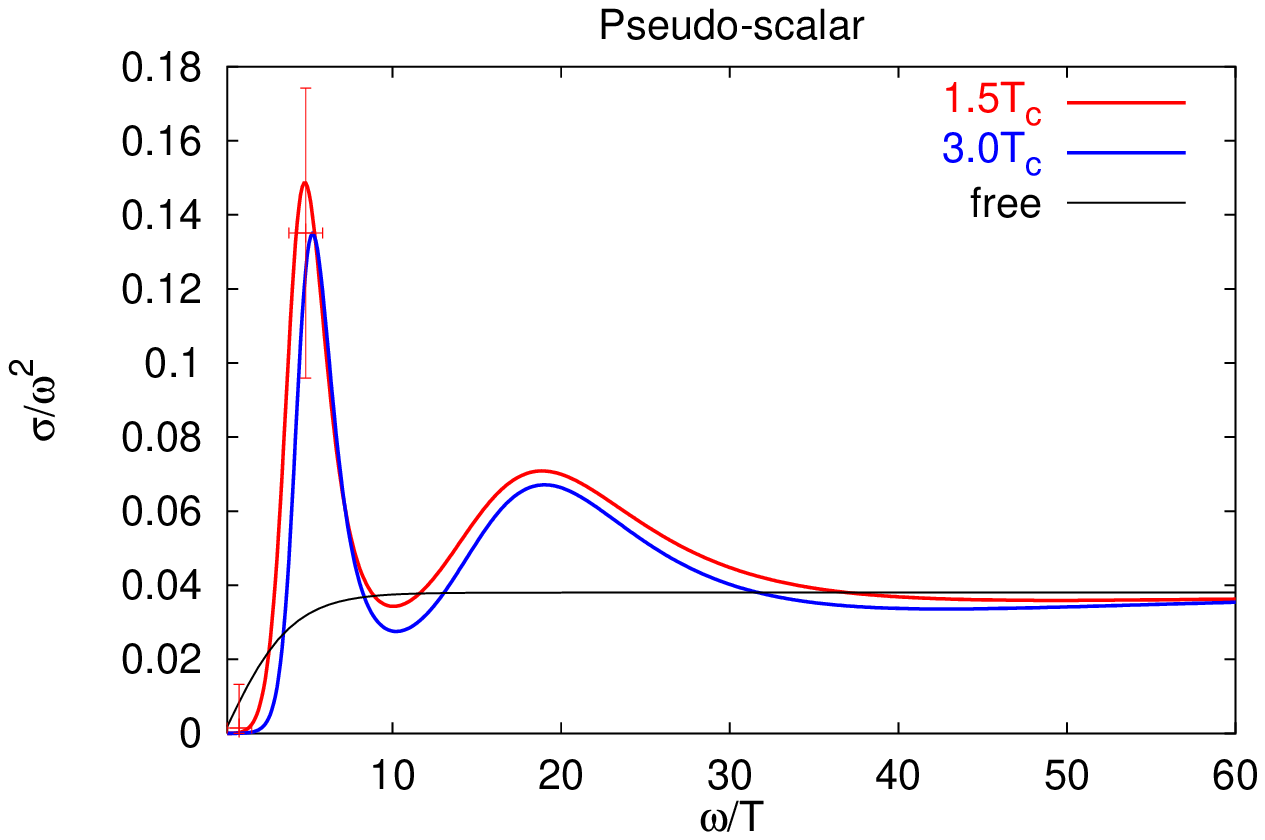}
\includegraphics[width=7cm]{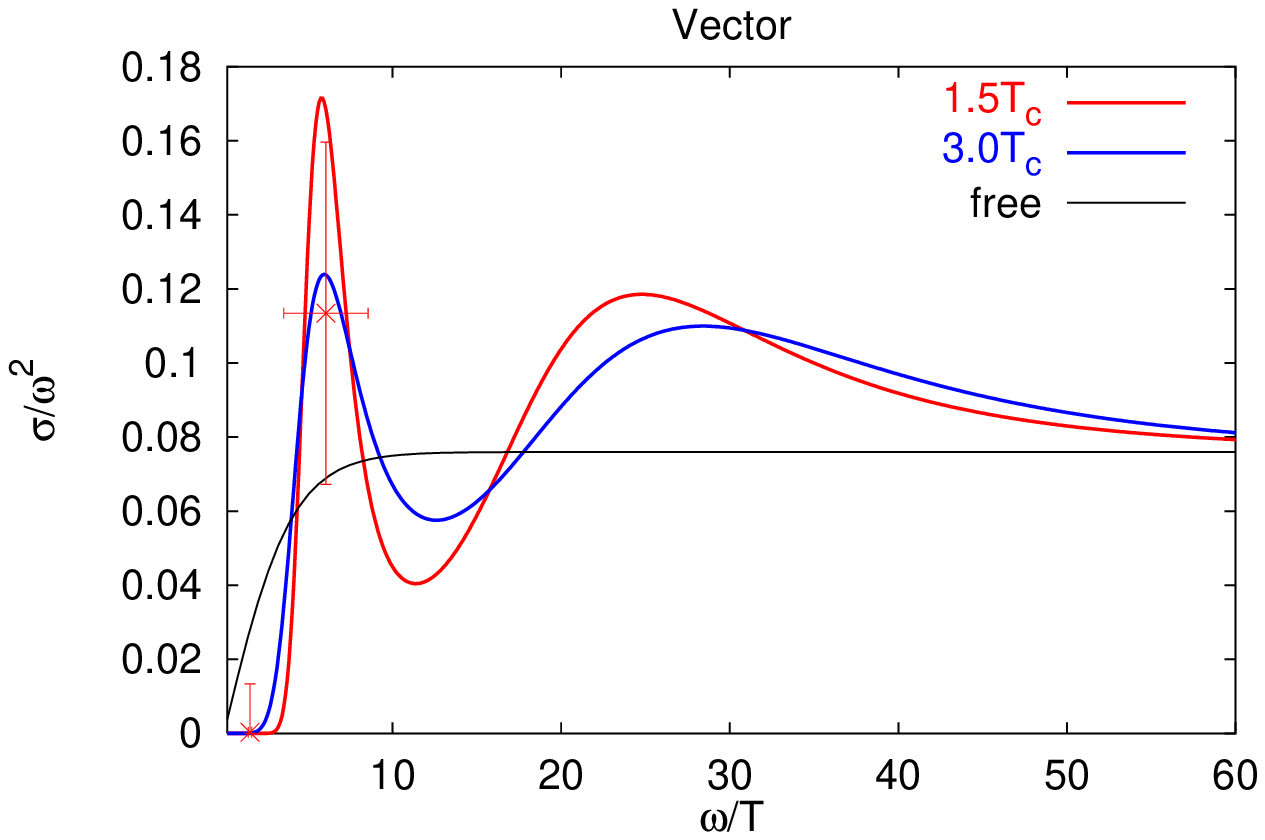}
\caption{The spectral functions in the deconfined phase for
the pseudoscalar channel (left) and the vector channel (right)
reconstructed using MEM on $64^3 \times 16$ lattice.}
\label{spfpsvc}
\end{figure}

\begin{figure}
\hspace*{-1cm}
\includegraphics[width=7cm]{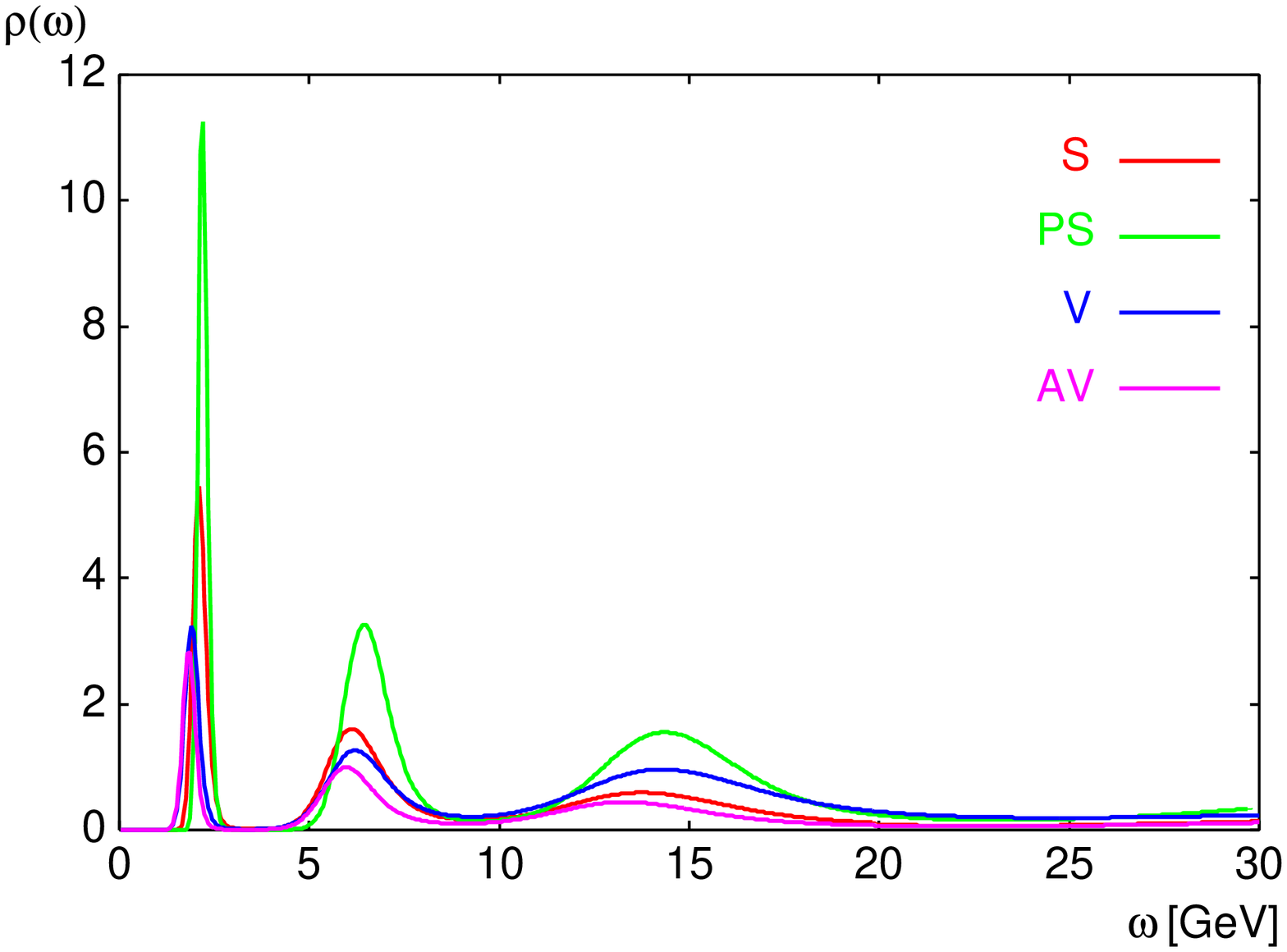}
\includegraphics[width=7cm]{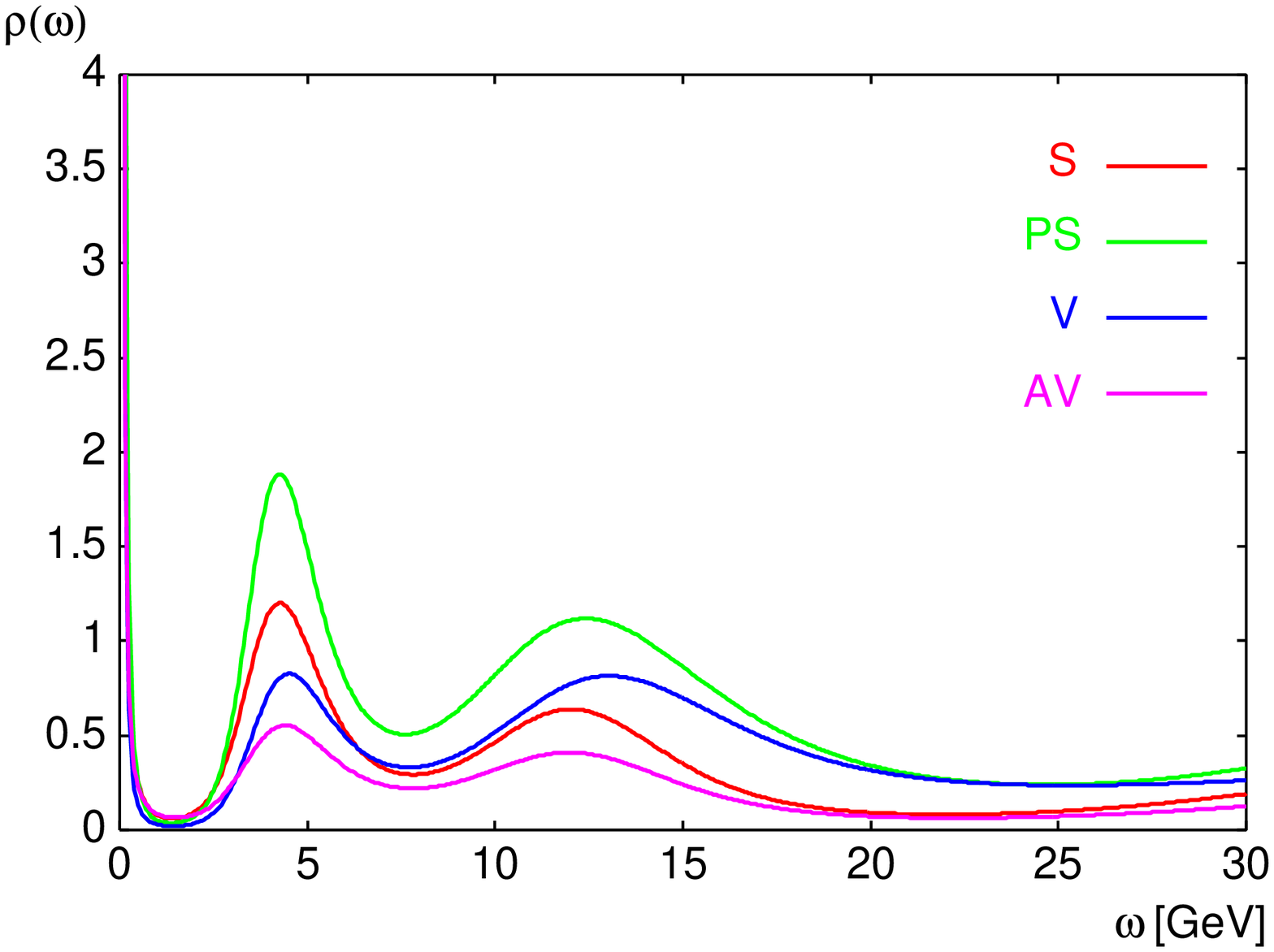}
\caption{The spectral functions $\rho(\omega)=\sigma(\omega)/\omega^2$ 
in different channels calculated 
at $1.4T_c$ (left) and at $1.9T_c$ (right) using anisotropic lattices \cite{asakawa01}
Note that normalization of the spectral functions is different from the case on
isotropic lattices.}
\label{asak}
\end{figure}

Before discussing the results on the spectral function 
above deconfinement let us first
look at the correlators. The correlation functions
alone can provide some constraints on the spectral functions,
for example the analysis of the vector correlation function 
provides stringent constraint on the thermal dilepton rate
\cite{karsch02}. At high temperature one would expect that the 
quark propagators should be close to the free ones. Therefore
in what follows we will always normalize  the meson
correlators by the corresponding lattice correlators in the free theory.
We will restrict the discussion to the case of zero spatial momentum $\vec{p}=0$.
In Fig. \ref{restor} I show the meson correlators at $1.5T_c$ in different
channels calculated on $64^3 \times 16$ lattice. As one can see from
the figure the scalar-pseudoscalar and vector-axial-vector correlators
become degenerate except at very small $\tau T$ where lattice artifacts
present in the Wilson formulation explicitly break chiral symmetry. 
This can be viewed as an indication of the $U_A(1)$ symmetry restoration at
high temperature. In Fig. \ref{corpsvc} I show the vector and pseudoscalar
correlators for $1.5T_c$ and $3T_c$ for different values of $N_{\tau}$.
From the figure it is evident that the vector correlator stays close
to its free value even at $1.5T_c$ while the pseudoscalar correlator is
considerably enhanced compared to the corresponding free correlators.
Note also that in the vector channel the correlators is $N_{\tau}$
(i.e. lattice spacing) independent, while  
in the pseudoscalar case there is a small $N_{\tau}$ dependence 
which is probably due to the perturbative error in calculation of 
$Z_{PS}$. In Fig. \ref{spfpsvc} I show the spectral functions
for the pseudoscalar and vector channels. The spectral functions
in both channels show a peak like structure at $\omega \simeq (5-6)T$
which is more pronounced in the
pseudoscalar case and probably leads to the large enhancement 
of the pseudoscalar correlator over the free case discussed above.
Note that the position of the peak appears to be proportional to the temperature.
Spectral function calculated on anisotropic lattice \cite{asakawa01} are shown in Fig. \ref{asak}
and exhibit  a similar peak like structure, 
roughly for the same value of $\omega/T$. However, at lower temperature, 
$T=1.4T_c$ the corresponding structure in the spectral functions appears
to be more sharp (see Fig. \ref{asak}).

\begin{figure}
\hspace*{-1cm}
\includegraphics[width=7cm]{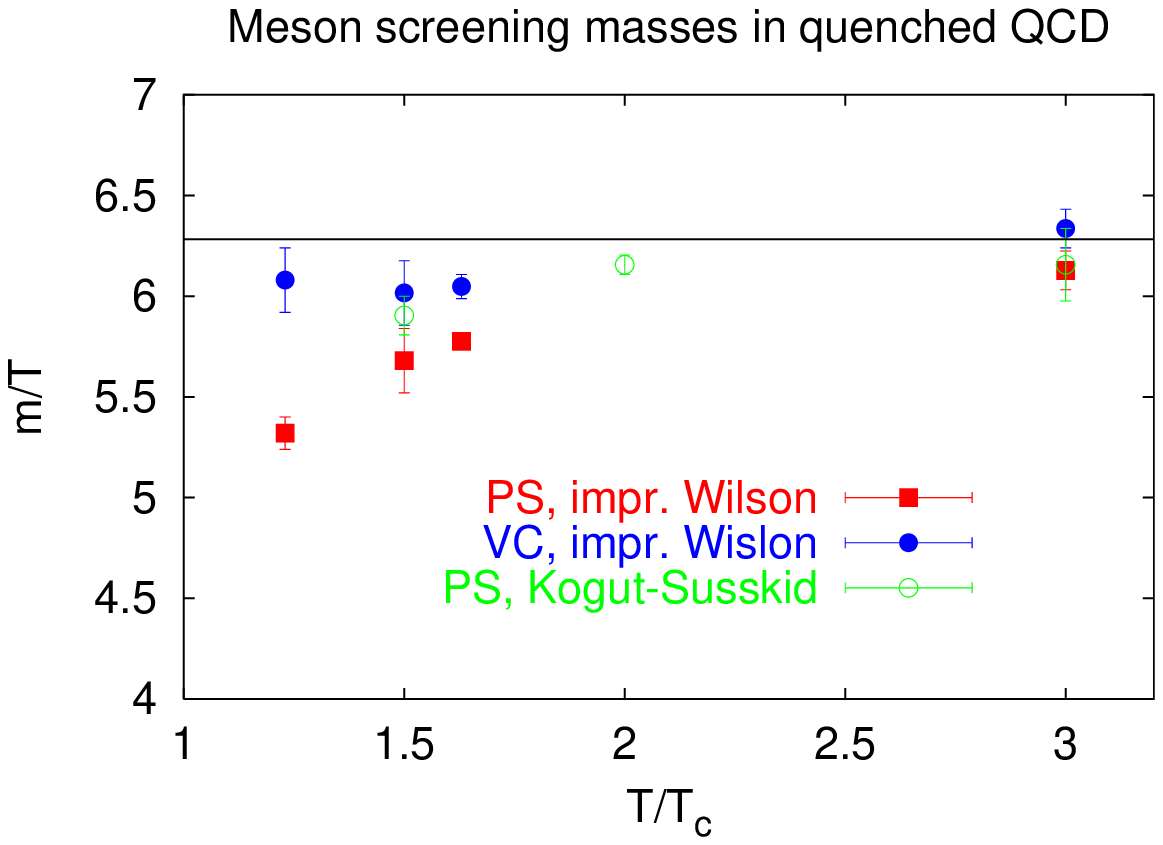}
\includegraphics[width=7cm]{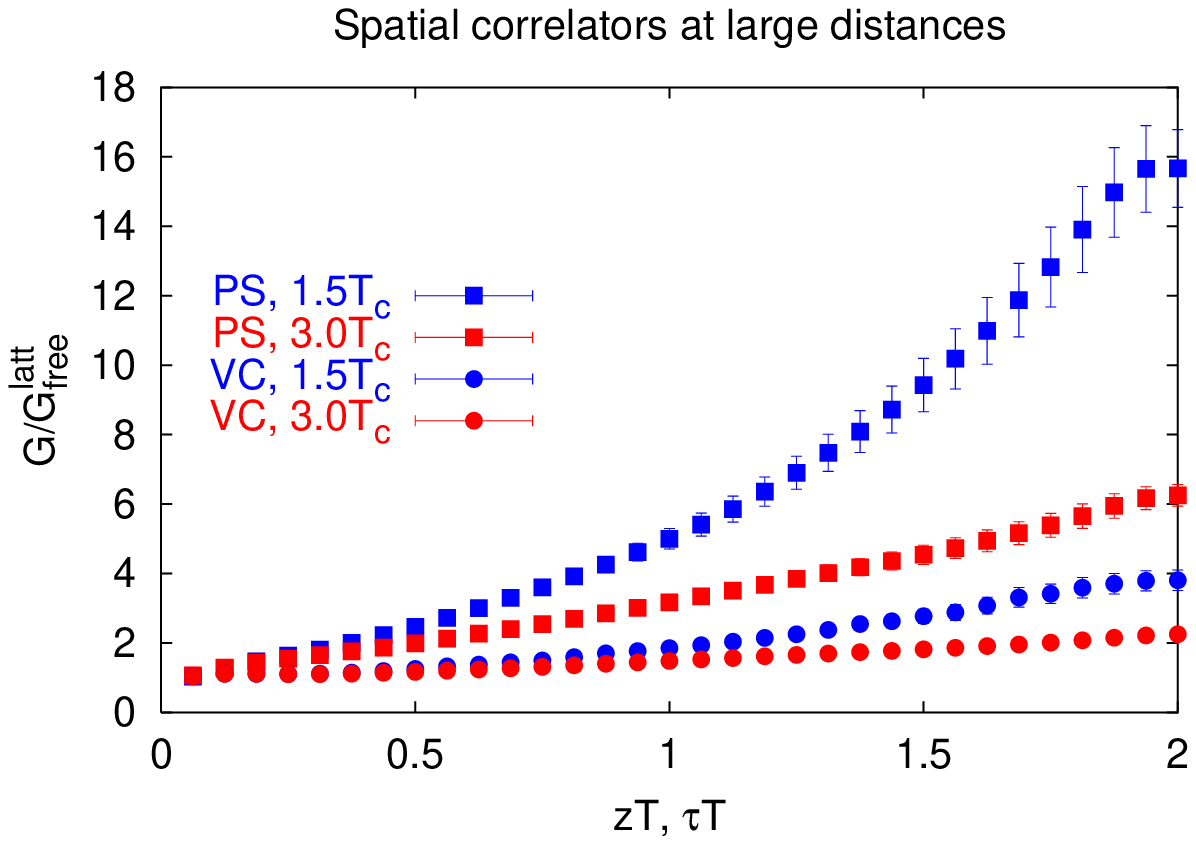}
\caption{The screening masses (left) and spatial correlators (right) for
vector and pseudoscalar channels at $T>T_c$. The screening masses have been
calculated in quenched QCD using improved Wilson fermions \cite{laermann,prog}
and Kogut-Susskid fermions \cite{gavai03}.}
\label{spat1}
\end{figure}

\begin{figure}
\hspace*{-1cm}
\includegraphics[width=7cm]{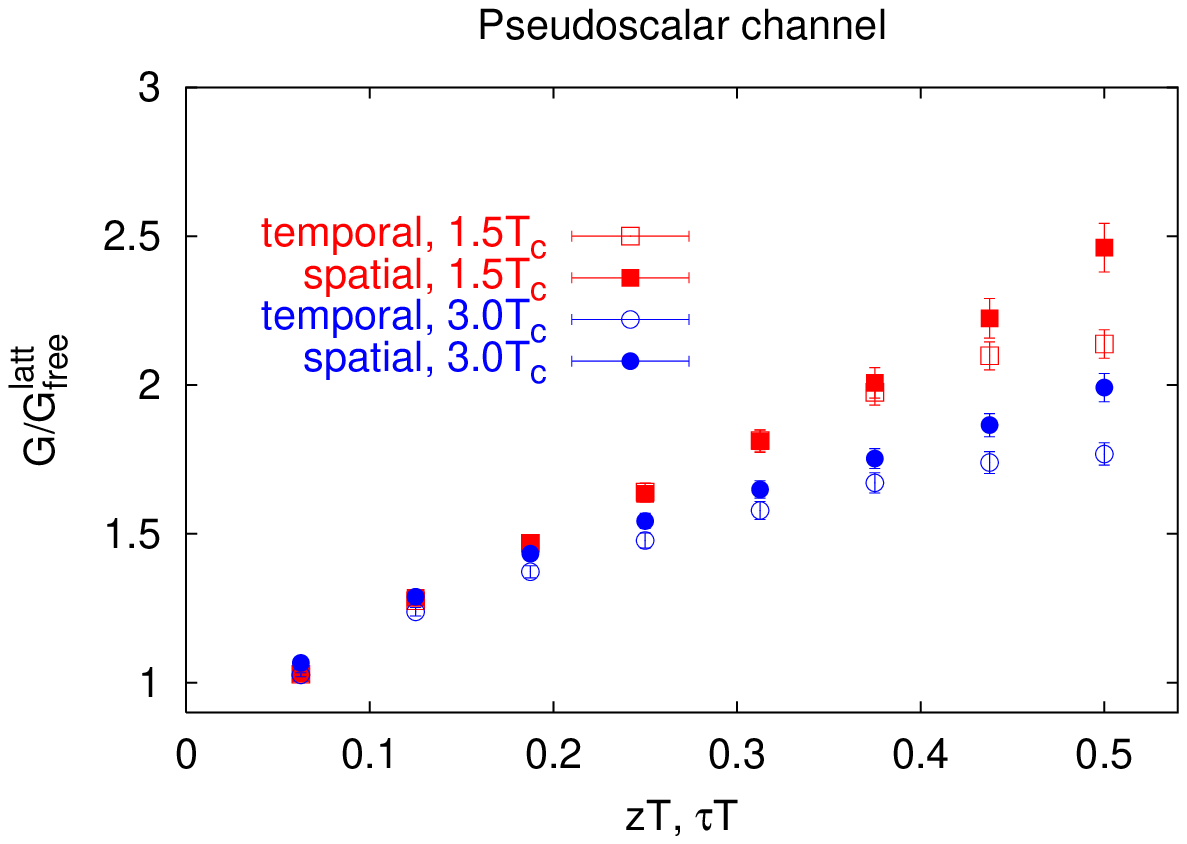}
\includegraphics[width=7cm]{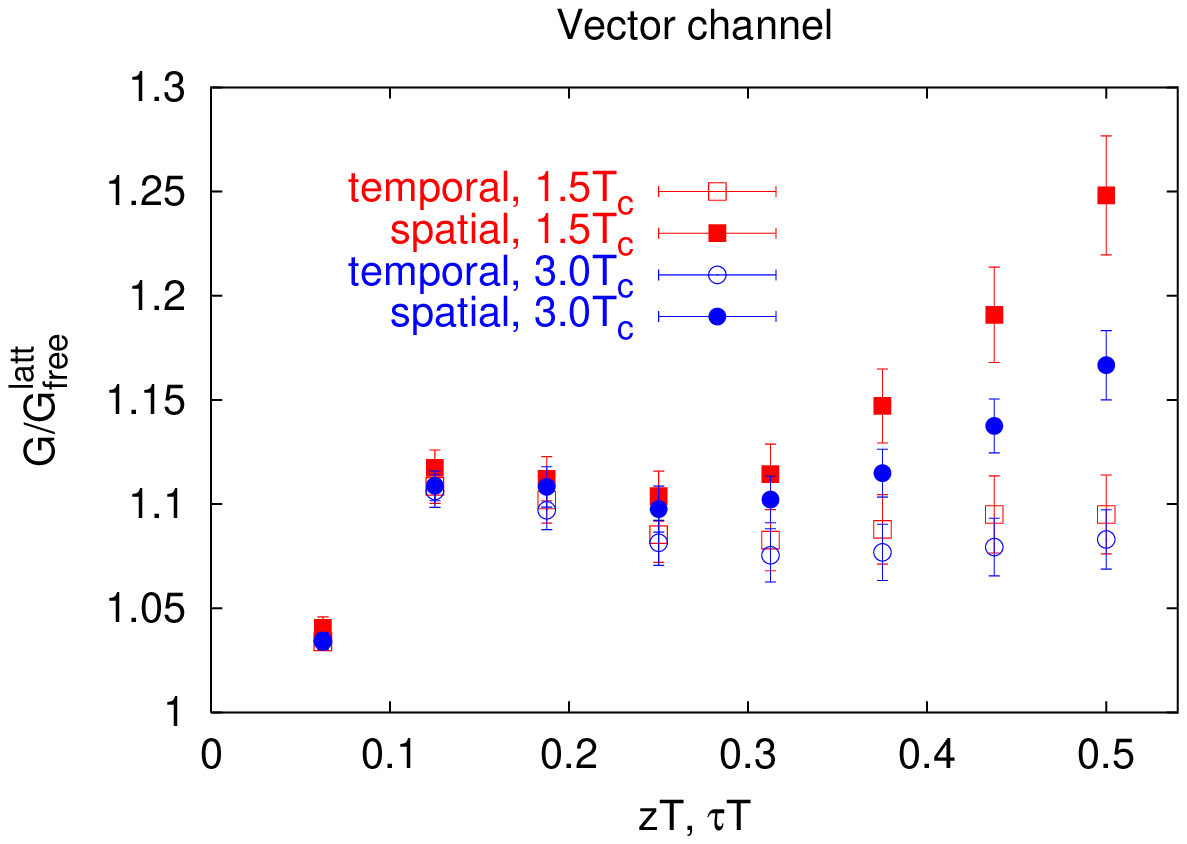}
\caption{
The ratio of spatial and temporal correlators in the
deconfined phase to the corresponding free correlators for
the pseudoscalar (left) and vector (right) channels.
}
\label{spat2}
\end{figure}

Meson correlators in the spatial directions and the corresponding
screening masses have been studied since long time
both in full and quenched QCD 
\cite{de86,de87,born91,bern92,gupta00,laermann,gupta-over,gavai03,prog}. 
However, reliable determination of
the screening masses is available only in quenched approximation 
\cite{laermann,gavai03,prog} as lattices with temporal extent $N_{\tau}\ge 8$
are needed in such analysis \cite{gavai03}. The most recent results 
for the screening masses are summarized in Fig. \ref{spat1}.
The screening masses are close to their asymptotic 
($T \rightarrow \infty $) value $2\pi T$ already at temperature for $T=1.5T_c$.
This could lead to the conclusion that there is an almost free quark propagation
in spatial directions in the deconfined phase at temperatures as low
as $1.5T_c$. However, a closer look on the spatial correlators in
Fig. \ref{spat1} reveals that this is not the case even at $3T_c$.
The correlators calculated by using lattice simulations are several times larger
than the corresponding free ones. This is not unexpected since it
is well known that the physics at distances $z>1/(g^2 T)$ 
is non-perturbative even at very high
temperature \cite{linde}.
In Fig. \ref{spat2} I show the vector and pseudoscalar spatial 
correlators at short distances together with the temporal one.
As one can see from the figures at small separations temporal
and spatial correlators show very similar behavior.

\section{Heavy quarkonia correlators and spectral functions}
\begin{figure}
\hspace*{-1cm}
\includegraphics[width=7cm]{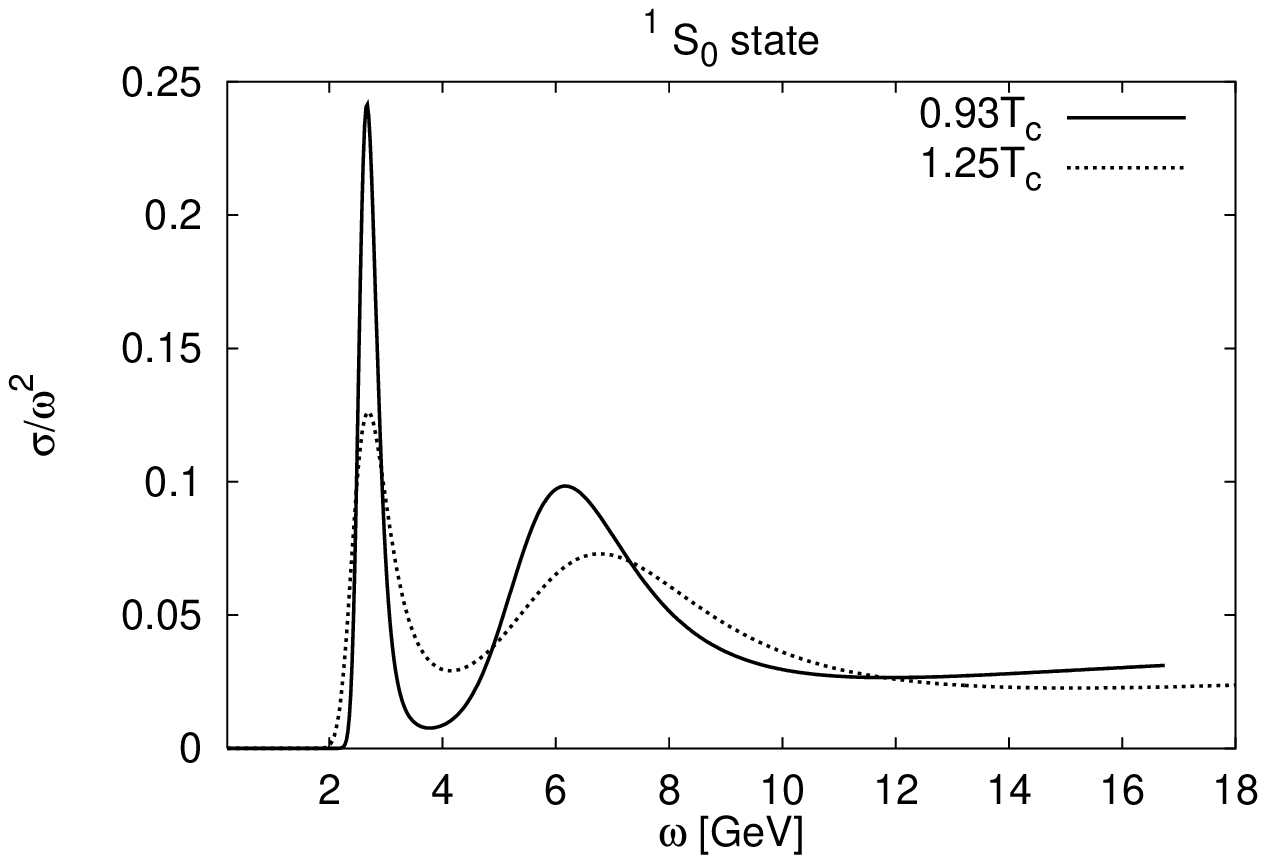}
\includegraphics[width=7cm]{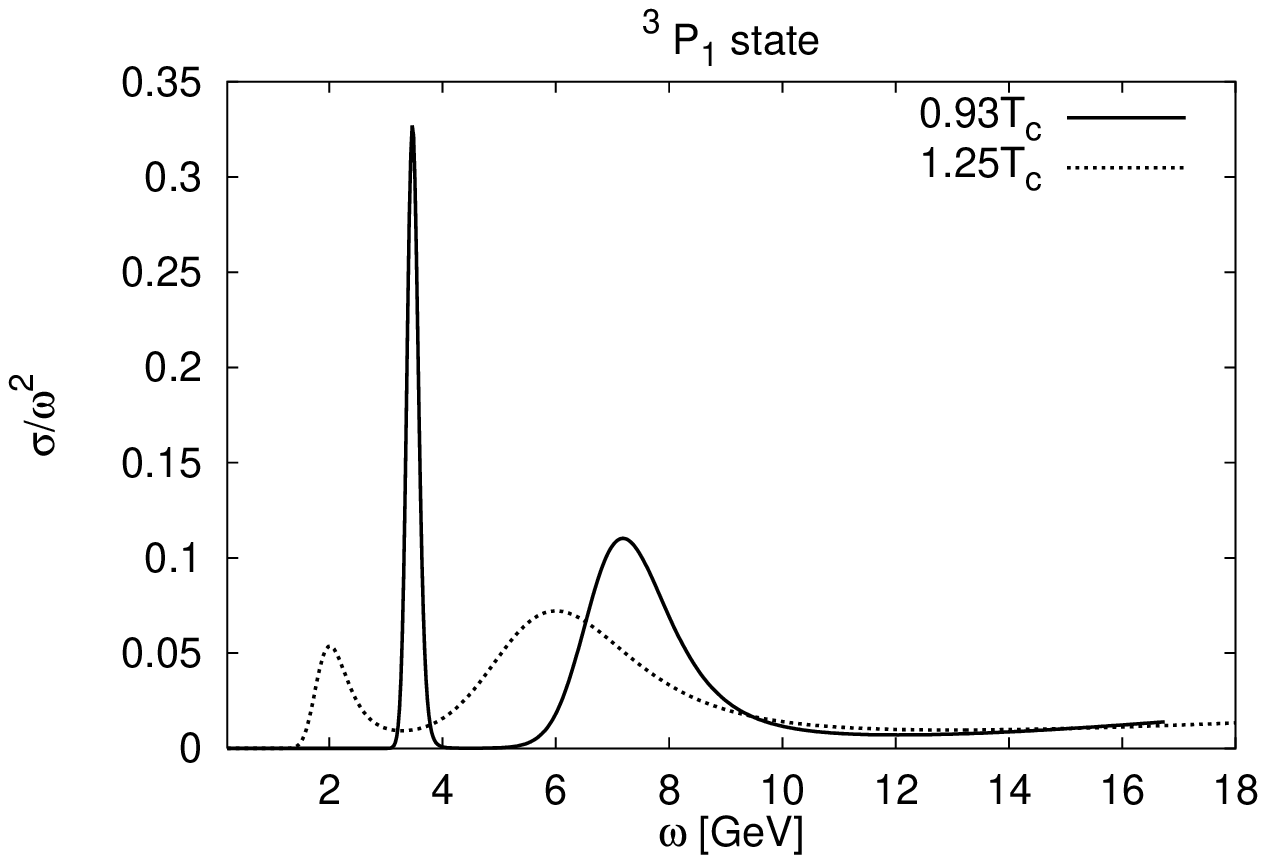}
\caption{Spectral function in the pseudoscalar
(left) and axial vector channels (right) reconstructed using MEM.
}
\label{heavyspf}
\end{figure}

\begin{figure}
\includegraphics[width=10cm]{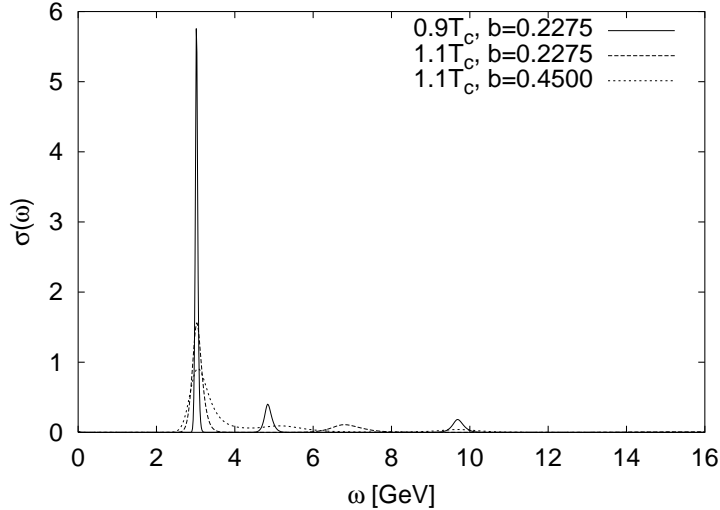}
\caption{
The $J/\psi$ spectral function reconstructed from the 
correlation function of extended operators using MEM \cite{umeda2}
for $0.9T_c$ and $1.1T_c$ and different input wave functions
(different values of $b$).
}
\label{umedafig1}
\end{figure}

Heavy quarkonia correlators and spectral functions were studied
in Refs. \cite{umeda2,umeda1} using anisotropic lattices
and extended operators as well as on isotropic lattices and point
correlators \cite{datta03}. In Ref. \cite{datta03} pseudoscalar, vector,
scalar and axial vector channels were considered which correspond
to $^1 S_0$ ($\eta_c$), $^3 S_1$ ($J/\psi$), $^3P_0$ ($\chi_{c0}$)
and $^3P_1$ ($\chi_{c1}$) charmonia states respectively.
As in the light quark sector some statements about in-medium
properties of charmonia can be made just by analyzing the 
behavior of the correlators. 
\begin{figure}
\hspace*{-1cm}
\includegraphics[width=7cm]{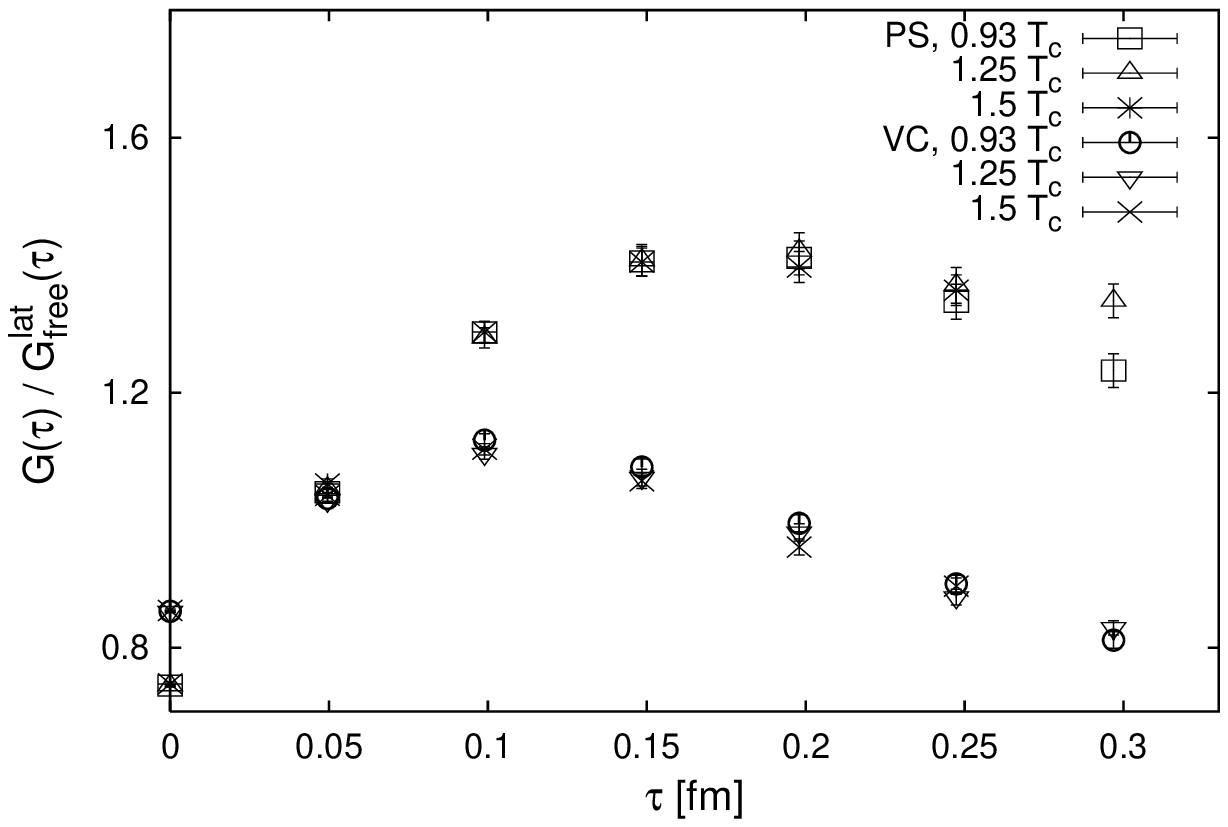}
\includegraphics[width=7cm]{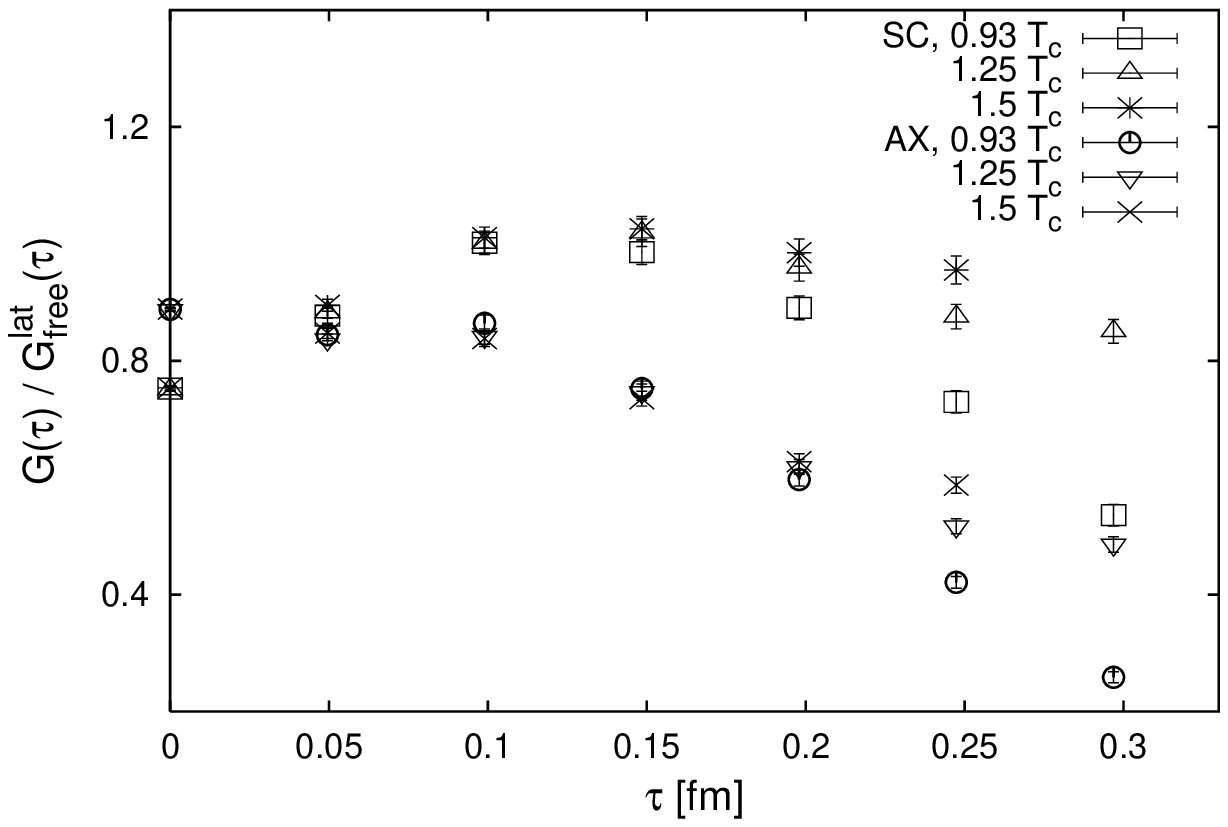}
\caption{The ratio of charmonia correlators to the 
corresponding free meson correlators in pseudoscalar and vector
channels (left) as well as for scalar and axial vector (right) channels
.}
\label{heavycor}
\end{figure}
In Fig. \ref{heavycor} I show 
the ratio of different quarkonia correlators to the corresponding
free correlators. As one can see from the figures the 
correlators in the pseudoscalar and vector channels show very little
change across $T_c$ while in the case of the scalar and axial vector
channels larger changes in the correlators are visible. This 
implies that the ground state charmonia ($^1 S_0$ and  $^3 S_1$)
are very likely to survive in the deconfined phase.
The spectral function reconstructed using MEM are shown in Fig. 
\ref{heavyspf} which shows indeed the the ground state peak
survives in the deconfined phase while the excited $^3P_0$ state
is dissociated already at $T=1.25T_c$. 
Spectral functions were also reconstructed for $1.5T_c$ and
and show again that the ground state charmonia survive while
the excited P-states are dissociated.
Ground state charmonia ( $^1 S_0$, $^3 S_1$ states) spectral functions 
were also studied in Ref. \cite{umeda2} using extended operators.
The results for the vector channel are shown in Fig. \ref{umedafig1}.
The $J/\psi$ mass (the peak position) changes very little across $T_c$
while the corresponding peak in the spectral function gets broader above
$T_c$. This broadening of the ground state peak is also visible 
in calculations with local operators. However, such broadening 
is not necessarily a physical effect and may come from the fact that
less data points $N_{\tau}$ are available at higher temperature
\cite{asakawa01} and also the temporal extent of the lattice
in physical units become smaller \cite{umeda2}. The authors
of Ref. \cite{umeda2} have found that the width of the peak
in the spectral function also depends on the trial wave function 
$\phi(y)$ used in constructing the extended operator (see Fig. \ref{umedafig1}),
the peak is broader for larger value of $b$ in the trial wave 
function $\phi(\vec{y})$. Therefore
MEM cannot yet provide a reliable estimate of the width of the
$J/\psi$ in the plasma. To overcome this difficulty $\chi^2$ -fits
to the Breit-Wigner form of the spectral function was used \cite{umeda2}.
This led to a first hints for a non-zero
widths of charmonium states above $T_c$.

The existence of ground state charmonia at $1.5T_c$ is in sharp
disagreement with predictions of potential models based on color 
screening \cite{mehr,digal01}. A possible reason for this
discrepancy is an oversimplified picture of color screening used
in these studies. In fact recent analysis of the free energy of 
quark-antiquark pair suggest quite complicated medium 
dependence of inter-quark forces \cite{okacz02,okacz03,digal03}.

\section{Conclusions}
In this paper recent results on meson spectral functions and
correlators calculated on lattice were discussed.
Though systematic uncertainties are present in the 
spectral functions the results appear to be very interesting and 
even intriguing to some extent.
The presence of the peak like structures in the meson spectral
functions well above deconfinement temperature is, too large
extent, unexpected. These peak structures seems to be present
in the spectral functions calculated on isotropic lattices 
\cite{karsch02,wetzorke02,datta03} as well as on anisotropic
lattices \cite{asakawa02,umeda2} and is some case their 
presence is supported by direct inspection of the corresponding
correlation functions (c.f. Fig. \ref{corpsvc} (left) and Fig. \ref{heavycor}).
In the case of heavy quarkonia the interpretation of the peaks in the
spectral function is clear, they correspond to the quarkonia states
which survive in the deconfined phase up to temperatures $1.5T_c$. 
We have seen that the position of these peaks does not change considerably
compared to the zero temperature case. The interpretation of peak-like
structures in the light quark sector is less evident as the position 
of these peaks seems to be proportional to the temperature. 
Certainly much more work is needed  in order to be able
to interpret the present findings in physical terms as well as 
for the detailed understanding of the systematic uncertainties involved in extraction of the
spectral functions from lattice correlators.

\vspace*{0.5cm}

\ack
Many results presented in the manuscript were obtained by
S. Datta, F. Karsch, E. Laermann, S. Stickan and I. Wetzorke and myself.
I would like to thank my colleagues for the collaboration. I would like
to thank S. Stickan and F. Karsch for careful reading of the manuscript. Finally
e-mail correspondence from T. Yamazaki, M. Asakawa and T. Umeda regarding
their results is gratefully acknowledged.


\vspace*{0.5cm}

\begin{thebibliography}{99}
\bibitem{rapp}
Rapp R, Wambach J 2000, {\em Adv. Nucl. Phys. } {\bf 25} 1;  Rapp R, Grandchamp L,  hep-ph/0305143
\bibitem{nakahara99}
Nakahara Y, Asakawa M and Hatsuda T 1999, \PR {\bf 60} 091503
\bibitem{asakawa01}
Asakawa M, Hatsuda T, Nakahara Y 2001, {\em Prog. Part. Nucl. Phys.} {\bf 46} 459
\bibitem{cppacs}
CP-PACS Collaboration, Yamazaki T 2001, \PR {\bf D65} 014501
\bibitem{karsch02}
Karsch F et al 2002, \PL {\bf B530} 147
\bibitem{wetzorke02}
Wetzorke I et al 2002, {\em Nucl. Phys. B (Proc. Suppl.)} {\bf 106} 513
\bibitem{karsch03}
Karsch F et al 2003, {\NP} {\bf A715} 701c
\bibitem{datta03}
Datta S et al 2003,  {\em Nucl. Phys. B (Proc. Suppl.) } {\bf 119} 487 (hep-lat/0208012)
\bibitem{asakawa02}
Asakawa Y, Hatsuda T, Nakahara Y, hep-lat/0208059
\bibitem{umeda2}
Umeda T, Nomura K, Matsufuru H, hep-lat/0211003 
\bibitem{wetzorke03}
Wetzorke I, hep-lat/0305012
\bibitem{lebellac}
Le Bellac M 1996, {\em Thermal Field Theory} ( Cambridge University Press )
\bibitem{lepage}
Lepage G P et al 2002, {\em Nucl. Phys. B ( Proc. Suppl.)} {\bf 106} 12
\bibitem{gupta-v}
Gupta S, hep-lat/0301006
\bibitem{umeda1}
Umeda T et al 2001, {\em Int. J. Mod. Phys. } {\bf A16} 2215
\bibitem{karsch03a}
Karsch F et al, hep-lat/0303017
\bibitem{biet96}
Bietenholz W et al 1997, 
{\em Nucl. Phys. B (Proc. Suppl.)} {\bf 53} 921
\bibitem{qcdtaro}
QCD-TARO Collaboration, de Forcrand Ph et al 2001, \PR {\bf D63} 054501
\bibitem{luescher}
L\"uscher M et al 1997, \NP {\bf B491} 344
\bibitem{de86}
DeGrand T A, DeTar C E 1986, \PR {\bf D34} 2469
\bibitem{de87}
DeTar C E, Kogut J B 1987, \PR {\bf D36} 2828
\bibitem{born91}
Born K D et al 1991, \PRL {\bf 67} 302
\bibitem{bern92}
Bernard C et al 1992, \PRL {\bf 68} 2125
\bibitem{laermann}
Laermann E, Schmidt P 2001, {\em Eur. Phys. J.} {\bf C20} 541
\bibitem{gupta00}
Gavai R V, Gupta S 2000, \PRL {\bf 85} 2068
\bibitem{gupta-over}
Gavai R V, Gupta S 2002, \PR {\bf D65} 094504
\bibitem{gavai03}
Gavai R, Gupta S 2003, \PR  {\bf D67} 034501
\bibitem{prog}
F Karsch et al, work in progress
\bibitem{linde}
Linde A 1980, \PL {\bf B96} 289
\bibitem{mehr}
Karsch F, Mehr M T, Satz H 1988, \ZP {\bf C37} 617
\bibitem{digal01}   
Digal S, Petreczky P, Satz H (2001), \PR {\bf D64} 094015 
\bibitem{okacz02}
Kaczmarek 0 et al 2002, \PL {\bf B543} 41 
\bibitem{okacz03}
Zantow F et al, hep-lat/0301015
\bibitem{digal03}
Digal S, Fortunato S, Petreczky P, hep-lat/0304017 
\end{thebibliography}
\end{document}